\documentclass{llncs}
\usepackage{hyperref}
\usepackage{amsmath}
\usepackage{graphicx}
\usepackage{caption}
\usepackage{amssymb}
\usepackage{subfig}
\usepackage{verbatim}
\usepackage{array}
\usepackage{multirow}
\usepackage{color}
\begin{document}
		\title{Tracking the Invisible: Privacy-Preserving Contact Tracing to Control the Spread of a Virus}

%	\title{Tracking and Controlling the Spread of a Virus in a Privacy-Preserving Way\thanks{An earlier version of this work is available at: \url{https://arxiv.org/pdf/2003.13073.pdf}}}

\author{Didem Demirag\inst{1} \and  Erman Ayday\inst{2,3}}
\institute{	Concordia University \\Montreal, QC, Canada \\ \email{	d\_demira@encs.concordia.ca}  \and 	Case Western Reserve University, \\Cleveland, OH, USA \\ \email{	exa208@case.edu} \and 	Bilkent University, Turkey}

\iffalse	
	\author{
		{\rm Didem Demirag}\\
		Concordia University\\
		Montreal, QC, Canada\\
		d\_demira@encs.concordia.ca
		\and
		{\rm Erman Ayday}\\
		Case Western Reserve University\\
		Cleveland, OH, USA\\
		and\\
		Bilkent University, Turkey\\
		exa208@case.edu
		% copy the following lines to add more authors
	} % end author
\fi	
	%\vspace{-5mm}
	\maketitle
\begin{abstract}
Today, tracking and controlling the spread of a virus is a crucial need for almost all countries. Doing this early would save millions of lives and help countries keep a stable economy. The easiest way to control the spread of a virus is to immediately inform the individuals who recently had close contact with the diagnosed patients. However, to achieve this, a centralized authority (e.g., a health authority) needs detailed location information from both healthy individuals and diagnosed patients. Thus, such an approach, although beneficial to control the spread of a virus, results in serious privacy concerns, and hence privacy-preserving solutions are required to solve this problem. Previous works on this topic either (i) compromise privacy (especially privacy of diagnosed patients) to have better efficiency or (ii) provide unscalable solutions. In this work, we propose a technique based on private set intersection between physical contact histories of individuals (that are recorded using smart phones) and a centralized database (run by a health authority) that keeps the identities of the positive diagnosed patients for the disease. Proposed solution protects the location privacy of both healthy individuals and diagnosed patients and it guarantees that the identities of the diagnosed patients remain hidden from other individuals. Notably, proposed scheme allows individuals to receive warning messages indicating their previous contacts with a positive diagnosed patient. Such warning messages will help them realize the risk and isolate themselves from other people. We make sure that the warning messages are only observed by the corresponding individuals and not by the health authority. We also implement the proposed scheme and show its efficiency and scalability via simulations.  
\end{abstract}
 
\section{Introduction}

A pandemic, which typically occurs due to uncontrollable spread of a virus, is a major threat for the mankind. It may have serious consequences including people losing their lives and economical devastation for countries. To decrease the severity of such consequences, it is crucial for countries to track the spread of a virus before it becomes widespread.

Main threat during such a spread is the individuals that had close contact with the carriers of the disease (i.e., people carrying the virus before they are diagnosed with the disease or before they start showing symptoms). Thus, it is very beneficial to identify and warn individuals that were in close contact with a carrier right after the carrier is diagnosed. 
If a country can identify who had close contact with the already diagnosed patients, by sending warnings to its citizens and telling them to self-quarantine themselves, the spread of the virus can be controlled. By doing so, individuals that receive such warnings (that they had close contact with one or more diagnosed patients) can take self-measures immediately. 
This is also economically preferred instead of completely shutting down a country.

However, implementing such an approach is not trivial due to privacy reasons. First of all, due to patient confidentiality, identities of diagnosed patients cannot be shared with other individuals. Similarly, healthy individuals do not want to share sensitive information about themselves (e.g., their whereabouts) with the authorities of the country. 

In this work, we propose a privacy-preserving technique that allows individuals receive warnings if they have been in close proximity of diagnosed patients in the past few weeks (that is determined based on the incubation period of the virus). We propose keeping the (physical) contact histories of individuals by using communication protocols in their smart phones. These contact histories are then used to determine if an individual was in close contact with a diagnosed patient in the past few weeks, and if so, the individual receives a warning. In order to do this in a privacy-preserving way, the proposed system uses private set intersection (PSI) on the background as the cryptographic building block (between the local contact histories of the individuals and database keeping the identities of the diagnosed patients). Even though PSI consumes more resources, it provides more privacy for the involving parties. By utilizing PSI, we aim to mitigate the privacy vulnerabilities of existing schemes, which will be explained in Section~\ref{sec:background}.

The proposed scheme guarantees that (i) identities or the whereabouts of the diagnosed patients are not revealed to any other individuals, (ii) contact histories of the individuals are not shared with any other parties, (iii) warning received by an individual (saying that they were in close proximity of a diagnosed patient) is only observed by the corresponding individual and no one else, and (iv) the individual that receives a warning can anonymously share their demographics with the healthcare officials only if they want to. Furthermore, we also propose an extension of the proposed scheme against malicious individuals that may try to tamper their local contact histories in order to learn the diagnosis of some target individuals. We also implement and evaluate the proposed technique to show its efficiency and practicality. 

The rest of the paper is organized as follows. In the next section, we summarize the related work. In Section~\ref{sec:background}, we provide brief background about the cryptographic building blocks we use in this paper. In Section~\ref{sec:solution}, we describe the proposed solution in detail. In Section~\ref{sec:evaluation}, we implement and evaluate the proposed scheme. In Section~\ref{sec:discussion}, we provide an extension of the proposed scheme and discuss its potential other uses. Finally, in Section~\ref{sec:conclusion}, we conclude the paper.

\section{Related Work}\label{sec:related}

The importance of outbreak and disease surveillance (without considering the privacy) has been studied by many previous works~\cite{bhatia2019big,carneiro2009google}. Privacy considerations are addressed in~\cite{cho2020contact}, where authors provide a study of existing contact tracing mobile apps for COVID-19 %in terms of their privacy considerations. 
Authors show that none of the existing apps and none of the existing schemes (except for private messaging systems) can protect the privacy of diagnosed patients and other exposed individuals at the same time. However, private messaging systems (in which, a diagnosed patient anonymously sends messages to its previous physical contacts) lack scalability and they heavily rely on proxy servers to obfuscate the identity of the diagnosed patient. Moreover, some countries prefer cell phone tracking-based systems (for diagnosed patients) in order to warn other individuals about the locations of diagnosed patients. However, such systems compromise privacy of individuals to track the spread of a virus~\cite{sciencemag}. 

Most of the current proposed systems are Bluetooth-based. Covid Watch ~\cite{covidwatch} is an open source project that relies on Bluetooth for contact tracing and also uses anonymized GPS data to detect high-risk areas. The server keeps contact event numbers from diagnosed users and users compare server's list with their contact list to find out whether they were in contact with a diagnosed person. In~\cite{singapore}, a health authority receives information from the diagnosed people and contacts people that were in contact with the diagnosed person. The health authority keeps record of phone numbers of users. Chan et al. offers functionalities to support the tracking of COVID-19, while also respecting security and privacy requirements by aiming to avoid using trusted third parties~\cite{chan2020pact}. They also discuss the risks of de-anonymization of a user. CONTAIN is another protocol that offers a privacy-preserving solution using Bluetooth~\cite{hekmati2020contain}. The system does not collect and log privacy-sensitive information. Furthermore, Epione~\cite{trieu2020epione} is a PSI-CA based system, where a new semi-honest PSI-CA primitive for asymmetric sets is used. While Epione is very similar to our proposed work, we published our initial idea~\cite{demirag2020tracking} on arxiv simultaneously with Epione.

A privacy-preserving Bluetooth protocol for contact tracing is proposed in~\cite{apple}. Here, each user owns a unique tracing key, from which the user can generate the keys needed for contact tracing. The system relies on Bluetooth for detecting the devices in the proximity. Users receive the information about diagnosed people from a diagnosis server and the matching is done locally on the user's device. 
In a similar system proposed in~\cite{epfl}, users keep their local contact histories by broadcasting their ephemeral, pseudo-random IDs from their smart phones and recording the IDs of other users that are in close proximity. A diagnosed individual voluntarily notifies a server and other users obtain this information from the server. Using this information from the server, the risk of a person for contracting the disease is computed locally on their phone. However, this system is not robust against a malicious user that may try to identify the infected individuals by observing (or modifying) their contact history.
%Also, while the authors prefer not to use privacy-preserving protocols like PSI due to considerations about scalability, we offer a solution based on PSI and show that it can scale even the input sizes from involving parties are significant.

Some apps, including~\cite{renprivacy,raskar2020apps,medium} track individuals' location and save it in a local database. However, such an approach compromises location privacy of diagnosed patients since upon diagnosis, location information is gathered by the health authorities. In~\cite{renprivacy,raskar2020apps,medium}, aggregate location paths (of diagnosed patients) are sent to the individuals and individuals get notification if they were in close proximity of a diagnosed patient. This results in another privacy vulnerability since individuals can observe the paths of diagnosed patients. Even though only the aggregate location paths are shared by the app, if the number of diagnosed patients is few in a given area, this may result in a significant privacy leak for the whereabouts of diagnosed individuals. As opposed to these approaches, in this work, we propose a scheme that protects the privacy of diagnosed individuals as well as the healthy ones. 

Some other works incorporate different building blocks as well. Liu et al. utilize a zero-knowledge protocol to protect privacy of the diagnosed patient and prevent false positive attacks, where either a patient who is not diagnosed with the disease pretend to be diagnosed or a diagnosed patient sends messages to people who are not their close contacts~\cite{liu2020privacy}. In~\cite{reichert2020privacy}, a secure multiparty computation-based approach is proposed. Another solution utilizes trusted execution environment (TEE)~\cite{enigma}. Here, users share encrypted location history and also the test status. The system uses private computation to determine the people who were in contact with a diagnosed person. Finally, in~\cite{xu2020beeptrace}, authors introduce BeepTrace, which is a blockchain-enabled approach.

%EA: last comment: what do you mean by power usage above? If you mean battery usage, we did not discuss about such a problem before.

%TO-DO

%[REF6]
%- voluntary to use the app
%- each device keeps its location history
%- once a patient is diagnosed with the disease, their location history is gathered by the server (not good for the privacy of diagnosed patients, they lose their location privacy)
%- location histories of diagnosed patients are aggregated to create the paths of infected patients
%- periodically such paths are downloaded by apps of individuals to see if they have been in the paths of infected patients (they only control location, not duration. Also they cannot do intersection of locations because locations will be in terms of x,y,z coordinates. So, they need to operate in plaintext in the local phone. If they do so, paths of infected patients can be learnt by other individuals. Furthermore, collection location history for 2 weeks is not scalable, especially if it is continuous tracking)

\section{Background}\label{sec:background}

Here, we briefly introduce the cryptographic building blocks we use in this paper.

\subsection{Private Set Intersection Cardinality (PSI-CA)}

Private set intersection cardinality (PSI-CA) aims to compute the cardinality of the intersection of two sets belonging to two parties, namely client and server, without disclosing either set to the other party~\cite{de2012fast}. At the end of the protocol between the client and server, the client learns only the size of the intersection. The only information that is leaked to the respective parties is the upper bounds for the size of the client and server's inputs. We provide the details of the PSI-CA algorithm within the proposed protocol (in Section~\ref{sec:test}).

\subsection{Authorized Private Set Intersection (APSI)}\label{sec:APSI}

Similar to PSI-CA, authorized private set intersection (APSI) also aims to compute the intersection of two sets (belonging to two parties) in a privacy-preserving way~\cite{de2009practical}. Different from PSI-CA, APSI requires the authorization of the client input first. Thus, client input is signed by a mutually trusted authority and client also provides these signatures as the input of the algorithm. We provide the details of the APSI algorithm when we introduce an extension of the proposed algorithm against a malicious individual that may tamper with their local contact history (in Section~\ref{sec:discussion}).

\section{Proposed Solution}\label{sec:solution}

In this section, we first introduce our system and threat models and then, we describe the proposed solution in detail.

\subsection{System Model}

The proposed system includes healthy (or not yet diagnosed) individuals with smart phones, diagnosed patients for the disease, and a health authority (e.g., ministry of health or NIH). Individuals interact with the database of the health authority. In the following, we will describe the proposed scheme assuming a single database for the health authority. For the sake of generality, one can also assume multiple local databases for the health authority (e.g., located in different geographical regions). 

The health authority keeps the identities of the diagnosed patients and it does not want this information to be learnt by other parties. Individuals keep their local (physical) contact histories in their smart phones and they do not want this information to be observed by other parties (including the health authority). Also, when an individual receives a warning about a contact with a diagnosed patient, the individual wants to make sure that no other party can observe this warning. 

\subsection{Threat Model}\label{sec:threat}

We consider a semi-honest attacker model for the parties that involve in the protocol. That is, each party in the system follows the protocol honestly but they may be curious to learn sensitive information of the other parties. On the other hand, individuals may try to learn the identities of diagnosed patients or the health authority may try to learn the contact histories of the individuals. As we will discuss in detail later, the proposed algorithm protects the parties against these threats. Finally, we assume all communications between parties (between smart phones of two individuals or between an individual and the database of the health authority) are encrypted, and hence robust against eavesdroppers.

In the following, we provide a list of possible attacks against the proposed system.
\begin{enumerate}
    \item A curious user trying to infer contact information or diagnosis belonging to a target user.
    \item A curious user trying to introduce fake contacts in the contact list,
    \item The server colluding with a curious user to infer contact information or diagnosis belonging to another target user.
\end{enumerate}

We discuss how these aforementioned attacks can be mitigated in Section~\ref{sec:mitigation}.

\iffalse
 \item leaking the identities of infected people (only the person and the health authority know)
 
from dp3t paper:
-eavesdropping the communication
-fake exposure events: making users think that they had a contact with an infected person even though they did not have any contact.
-suppressing at risk contacts: preventing users from learning the fact that they are exposed to the disease and they are at risk. 
-prevent contact discovery: this is related to jamming Bluetooth signals.
-``retroactive attacks": reidentifying indivuduals using the data stored on the device + information about the user's whereabouts
identifying locations that are visited by diagnosed people

the privacy questions seemed relevant (https://arxiv.org/pdf/2005.11416.pdf)
``Could external parties exploit your system to track users or infer whether they are infected?" 
\fi

\subsection{Keeping the Contact History at Local Devices}\label{sec:contact}

\begin{figure}[ht]
\centering
		\includegraphics[width=0.65\columnwidth]{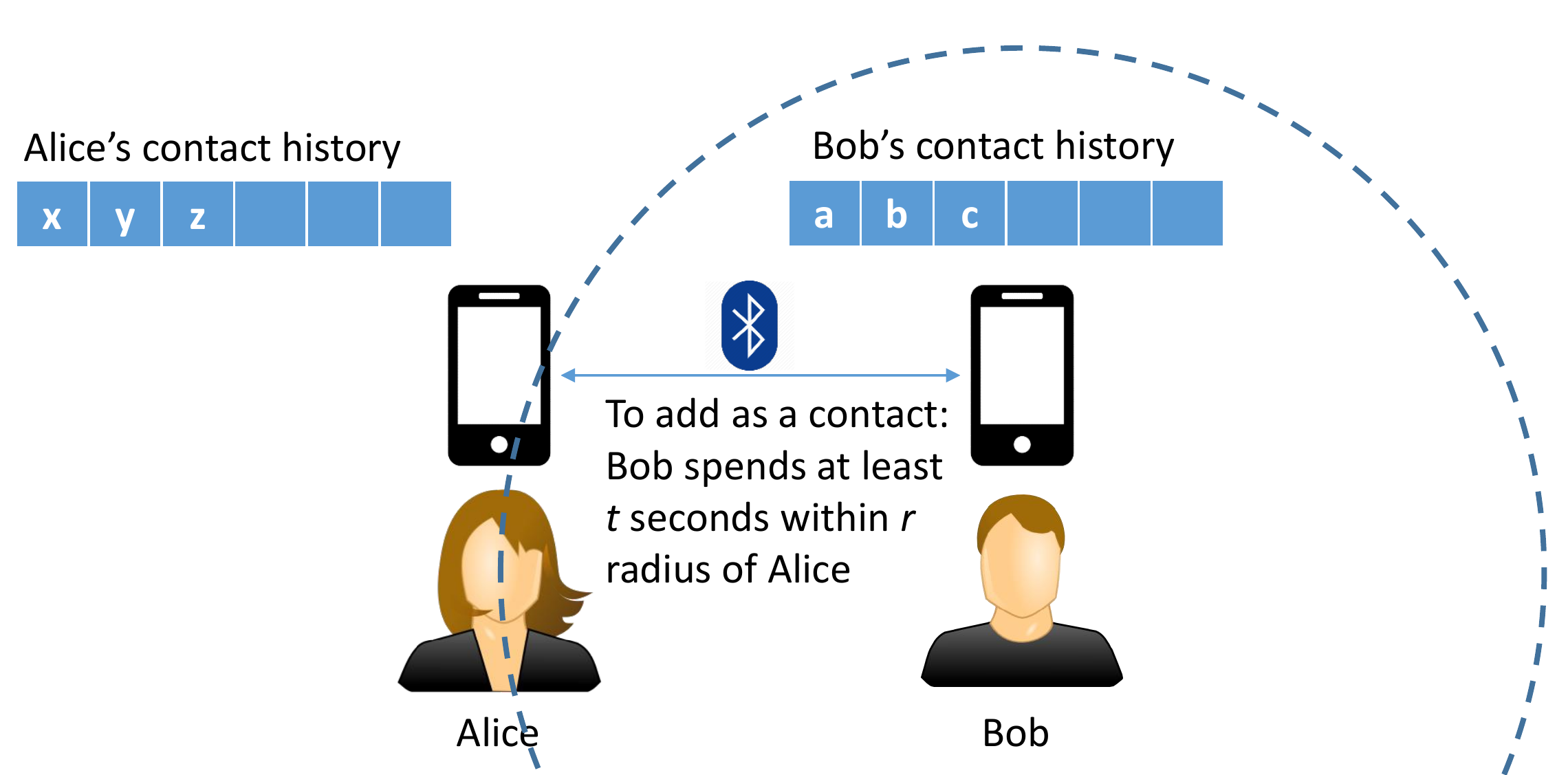}
		%\vspace{-5mm}
		\caption{Keeping and updating local contact histories of the individuals.}
		\label{fig:model}
		%\vspace{-3mm}
\end{figure}

Each individual keeps a vector in their local smart phone for their physical contact history. When an individual $A$ spends some amount of time within the close proximity of an individual $B$, their phone records the ID of person $B$. For IDs of the individuals, we propose using the hash of their User ID (UID), which is generated by the application. %For IDs of the individuals, we propose using the hash of their IMEI or phone numbers.
To measure the proximity between the individuals, we propose using the Bluetooth signals on their devices. Thus, to add individual $A$ as a contact, $B$ needs to spend at least $t$ seconds within $r$ radius of $A$. This process is also illustrated in Figure~\ref{fig:model}.
As a result of this interaction, the new record in the contact history of $A$ is the ID of $B$. Each individual may also separately keep the time and the duration of the contact (to have more insight about their risk, as will be discussed later).
%The same record is created in contact histories of both $A$ and $B$. Note that location information in the contact record  optional and it may include the GPS coordinates of the individuals (a common location for the records of both individuals).

It has been shown that the strengths of Bluetooth signals can be used to approximate the distance between two devices~\cite{jung2013distance,zhou2006position}. Alternatively, one can also use (i) just the Bluetooth coverage (e.g., when two devices are in the range of each other for more than $t$ seconds, they can update their local contact histories with each others' IDs), (ii) GPS information (when two devices are within their Bluetooth coverage, they can exchange GPS information to measure their distance more accurately and update their local contacts if they spend more than $t$ seconds within a close range of each other), or (iii) NFC signal coverage (when the devices are within NFC signal coverage of each other, which is about 3 feet maximum, for more than $t$ seconds, they can update their local contact histories with each others' IDs). Since this part is not the main contribution of the paper, we do not go into the details of establishing the contact histories. 

It is important to make sure that an individual cannot add a contact in their contact list aiming to learn whether a target person has %positive diagnosis or not. For instance, knowing the IMEI number of a
positive diagnosis or not. For instance, knowing the UID of a
target person, an attacker may construct its local contact list only from the ID of that target, and hence learn the diagnosis of the target. To prevent such an attack, we consider two options: (i) make sure the local contact histories of individuals are stored in such a way that data cannot be accessed or modified by the individuals (e.g., the local contact history can be encrypted in the device by the key of the health authority or the contact history can be stored in a storage for which the individual does not have read/write permission). Or, (ii) each new contact $B$ of an individual $A$ also includes a digital signature that is signed by a centralized authority (e.g., the telecom operator). To do so, if individuals $A$ and $B$ spend a certain amount of time within close proximity of each other (measured as discussed before), they both send the contact request to the operator, the operator signs and sends back the signed contact record to both parties, and each party keeps the contact records and the corresponding signature together. This way, an attacker cannot fake new contacts in its local contact history. The validity of these signatures are then verified when the local contact history of an individual is compared with the diagnosed patients in the health authority's database (we discuss this in detail in Section~\ref{sec:discussion}).

Using either of these techniques, the developed algorithm makes sure that the contact history cannot be tampered by the individual. Note that if the storage of the local contact list would be an overhead, it is also possible to store the local contacts of an individual at a cloud server (encrypted by individual's key) and update the local contacts periodically. Current Bluetooth based solutions, such as~\cite{apple} do not let the users access their contact histories (that is stored on their mobile devices). Even if the user can access their contact history, tampering with this list can be avoided by using digital signatures and encryption, as explained in this section.

\subsection{Keeping the IDs of Diagnosed Patients at a Centralized Database}

\begin{figure}[ht]
\centering
		\includegraphics[width=0.55\columnwidth]{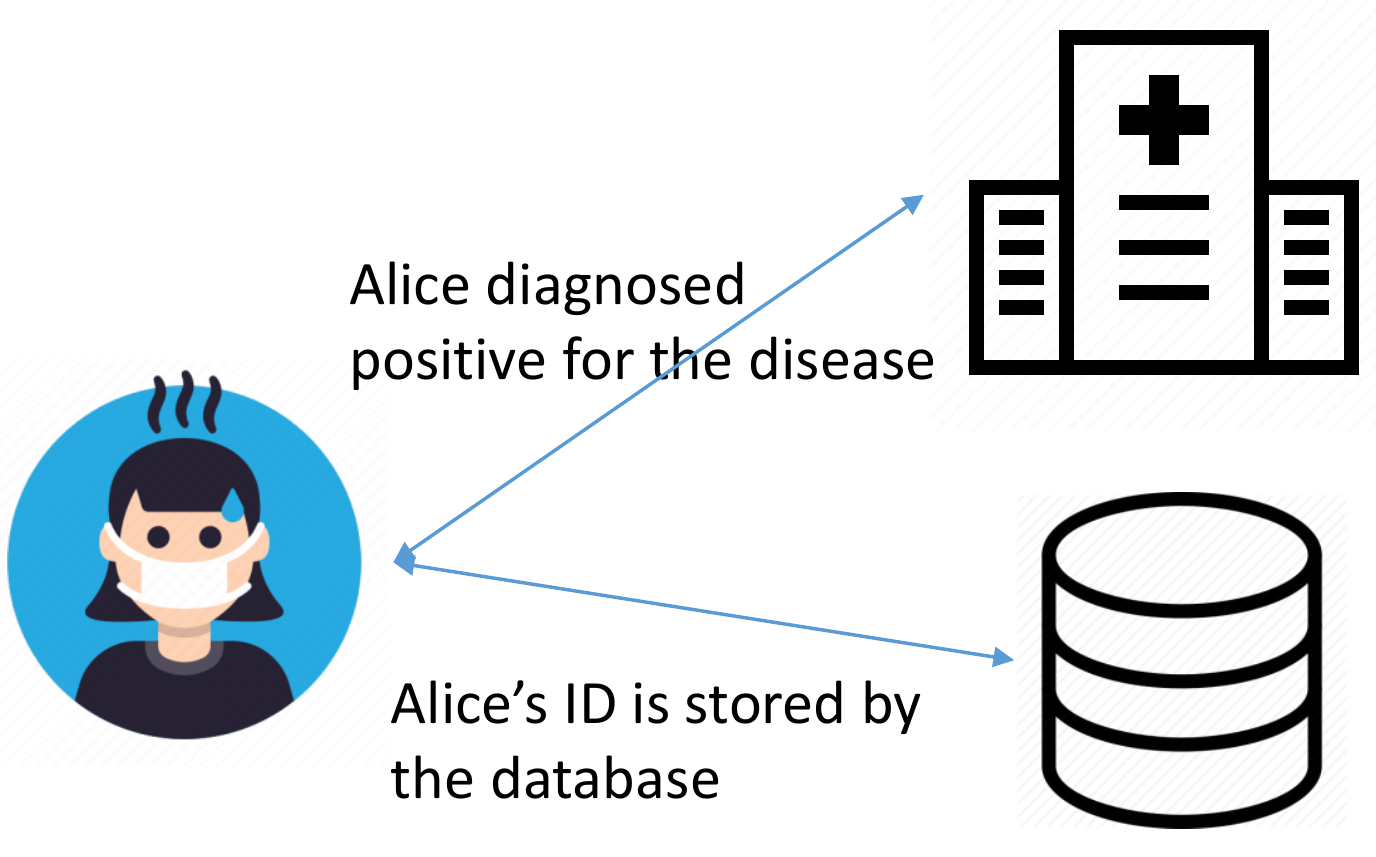}
		%\vspace{-5mm}
		\caption{Updating the database of the health authority with the IDs of the diagnosed patients.}
		\label{fig:diagnosis}
		%\vspace{-3mm}
\end{figure}

When an individual is diagnosed (e.g., by a hospital) with the %disease, the ID (hash of IMEI or phone number) of the positive
disease, the User ID (UID) of the positive diagnosed patient is stored in the database of the health authority (e.g., ministry of health), as shown in Figure~\ref{fig:diagnosis}. It is important to note that only the hospital and the health authority know the ID of the diagnosed individual. 

\subsection{Private Set Intersection to Identify the Individuals at Risk}\label{sec:test}

The application on an individual's smart phone sends queries to the health authority's database following a random schedule. This schedule can be determined by the system to avoid an overload to the database. It is also important not to allow the individual to send queries at any time in order to control the system's bandwidth. 

\begin{figure}[ht]
\centering
		\includegraphics[width=0.65\columnwidth]{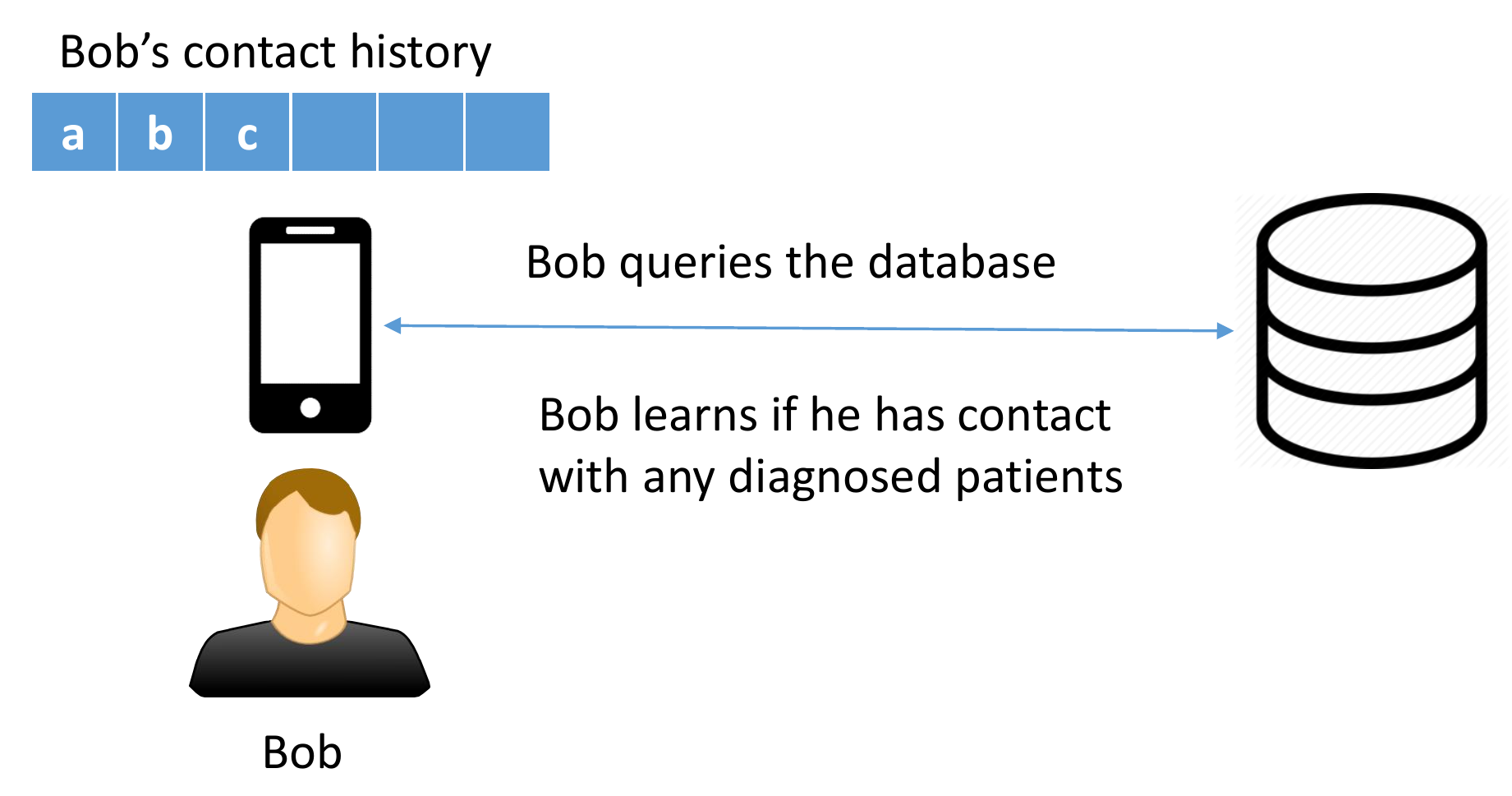}
		%\vspace{-5mm}
		\caption{Privacy-preserving interaction between an individual and the database of the health authority.}
		\label{fig:test}
		%\vspace{-3mm}
\end{figure}

An individual $A$ uses their contact history to query the database of the health authority and the goal is to identify whether there is an intersection between the local contact history of the individual and the IDs of the diagnosed patients in the authority's database (as shown in Figure~\ref{fig:test}). Size of this intersection reveals the number of diagnosed people with whom $A$ had been in close proximity in the previous a few weeks. As the size of the intersection increases, the risk of individual $A$ being infected also increases. The proposed algorithm provides the result of this intersection to individual $A$ as a warning message. Using the warning, the individual may take early precautions (e.g., have a test or quarantine themselves).

To compute this intersection in a privacy-preserving way, we use the private set intersection cardinality (PSI-CA) protocol, in which parties that are involved in the protocol obfuscate their inputs (sensitive information) and compute the result of this intersection. Eventually, only individual $A$ learns the result of the intersection and the health authority does not learn any information about the contact history of individual $A$ or the result of the intersection. We also make sure that individual $A$ does not learn anything about the database content of the health authority (e.g., IDs of diagnosed patients). We provide the details of this protocol in the following. 

\begin{figure}[ht]
\centering
		\includegraphics[width=0.80\columnwidth]{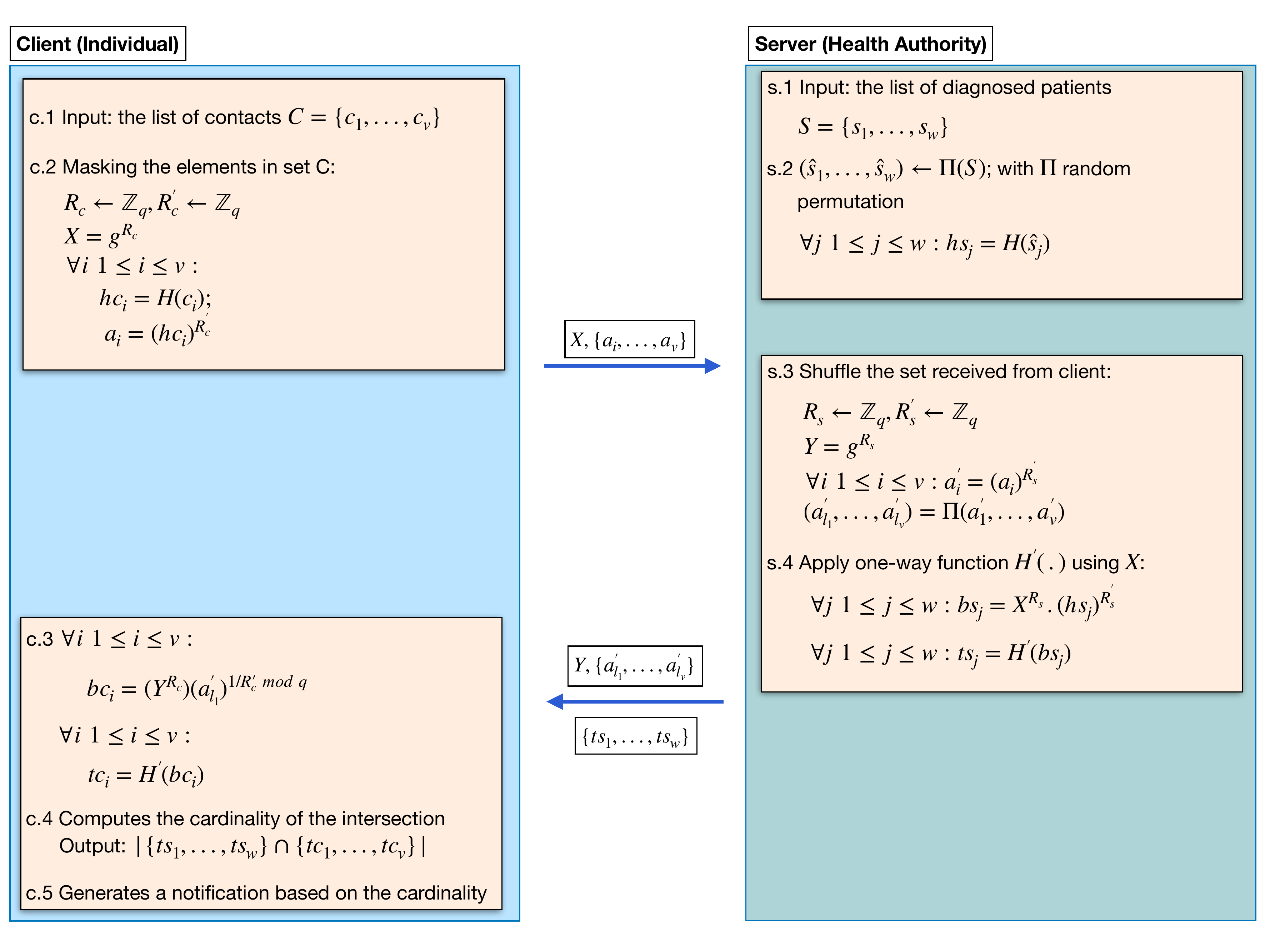}
		%\vspace{-5mm} %0.85
		\caption{Details of the PSI-CA based protocol between an individual (client) and the health authority (server).}
		\label{fig:PSI-CA}
		%\vspace{-3mm}
\end{figure}

Figure~\ref{fig:PSI-CA} illustrates the details of the proposed PSI-CA based protocol between an individual (client) and the health authority (server). As input to the protocol, client has its local contact list and server has the list of positive diagnosed patients (steps c.1 and s.1 in the figure). 

Client masks its input with the random exponent $R^{'}_c$ and obtains the list of $a_i$-s and computes $X=g^{R_c}$ ($X$ is similar to an ElGamal public key). Client sends the list of $a_i$ values and $X$ to the server (step c.2 in the figure). 

Server permutes its input list and applies $H(.)$ on the list (step s.2 in the figure). Server masks $a_i$ values with its random exponent $R^{'}_s$, shuffles the resulting list, and computes $Y=g^{R_s}$, which is a public-key like value (step s.3 in the figure). Server creates the list of $ts_j$-s by applying the one-way function $H(.)$ over the multiplication of $X^{R_s}$ and exponentiation of $hs_j$-s to random value $R_s^{'}$ (step s.4 in the figure). Server sends shuffled and masked $a_i$ values, $Y$, and $ts_j$-s to the client. 

As the last step, client does the matching between the list of $ts_j$-s that it received from the server and its own list of $tc_i$-s (step c.4 in the figure). $tc_i$-s are obtained by applying the one-way function $H(.)$ over the multiplication of $Y^{R_s}$ and the shuffled $a_i$-s, which are stripped of the random value $R_c^{'}$ (step c.3 in the figure). At the end of the protocol, client only learns the cardinality of the intersection. At step c.5 in the figure, a notification is generated based on the output of PSI-CA.

\subsection{Further Steps to Track the Spread}

Once an individual receives a warning as a result of the proposed algorithm, they can choose to (i) provide (anonymous) information back to the health authority to help the authority to track the spread and/or (ii) share their local contact history with the health authority to get further information about their risk. 

In (i), the individual shares their demographics, the size of intersection they obtain as a result of the proposed algorithm, and their location with the health authority. Using such information received from different individuals, the health authority can have a clear idea about how the virus spreads in the population. In (ii), the health authority, using the local contact history of the individual and further details about the contacts of individual (duration and location of each contact), can provide a more detailed risk information to the individual (e.g., if the duration of the contact with a diagnosed patient is long, then the risk of individual also increases). As discussed before, such contact details (i.e., duration or location) are not used in the proposed privacy-preserving algorithm; they can be collected and kept by the individual and may be shared with the health authority to get more insight about the risk.

\section{Evaluation}\label{sec:evaluation}

We implemented and evaluated the proposed privacy-preserving search algorithm. 
We ran our experiments on macOS High Sierra, 2.3 GHz Intel Core i5, 8GB RAM, and 256GB hard disk.
We used MD5 as the hash function to hash the UIDs of individuals in the local contact lists and in the health authority's database.
We used the implementation of PSI-CA in~\cite{de2012fast}, in which $q$ and $p$ are $160$ and $1024$ bits, respectively. We ran each experiment for $20$ times and reported the average.

We show the results of the evaluation of PSI-based solution in Tables~\ref{table:server} and~\ref{table:client}. In Table~\ref{table:server}, we set the size of client's local contact list to $1,000$ and vary the size of server's database. In Table~\ref{table:client}, we set the size of server's database to $100,000$ and vary the size of client's local contact list.
Our results show that the online phase of the protocol can be efficiently completed by the parties even when the input sizes of both parties are significantly large. 
Note that the server does not need to run the offline part of the algorithm for each client separately. Instead, the server can use the same offline computation during its interaction with every client. Also, a client can conduct its offline steps as it generates its local contact list.

Complexity of the proposed algorithm is linear in the size of the two sets. Let the cardinality of server's set be $w$ and client's set be $v$. Client performs $2(v+ 1)$ exponentiations with $|q|$-bit (short) exponents modulo $|p|$-bit and $v$ modular multiplications. Server performs $(v + w)$ modular exponentiations with short exponents and $w$ modular multiplications. The resulting communication overhead is $2(v + 1)$ $|p|$-bit and $w$ $\kappa$-bit values, where $\kappa$ is the security parameter. It can be deduced from these results that the protocol does not incur a significant overhead for a smart phone. Security and correctness of the proposed algorithm depends on the security and correctness of the original PSI-CA algorithm. We refer to~\cite{de2012fast} for details.

\begin{table*}[ht]
\caption{Offline and online run-times for PSI-based protocol (in milliseconds) at the client (individual) and server (health authority) with varying size for server's database. Size of client's contact list is set to $1,000$.}
\begin{center}
\begin{tabular}{ |c|c|c| } 
\hline
Size of server's database & Offline Time (ms) & Online Time (ms) \\
\hline
\multirow{2}{4em}{1,000} & Client: 210.6 & Client: 100.85 \\ 
& Server: 388.05  & Server: 107.5 \\ 
\hline
\multirow{2}{4em}{10,000} & Client: 201.2 & Client: 978.7\\ 
& Server: 2213.5  & Server: 1003.8 \\ 
\hline
\multirow{2}{4em}{100,000} & Client: 202.95 & Client: 9766.1 \\ 
& Server: 20054.1 & Server: 9925.6\\ 
\hline
\multirow{2}{4em}{1,000,000} & Client: 202.4 & Client: 96685.8 \\ 
& Server: 194289.6  & Server: 98631.1  \\ 
\hline
\end{tabular}
\end{center}
\label{table:server}
\end{table*}

\iffalse
\begin{table*}[ht]
\caption{Offline and online run-times (in milliseconds) at the client (individual) and server (health authority) with varying size for server's database. Size of client's contact list is set to $1,000$.}
\begin{center}
\begin{tabular}{ |c|c|c| } 
\hline
Size of server's database & Offline Time (ms) & Online Time (ms) \\
\hline
\multirow{2}{4em}{1,000} & Client: 210 & Client: 106 \\ 
& Server: 429 & Server: 104\\ 
\hline
\multirow{2}{4em}{10,000} & Client: 199 & Client:956 \\ 
& Server: 2136 & Server: 984\\ 
\hline
\multirow{2}{4em}{100,000} & Client: 203& Client: 9624 \\ 
& Server:20139 & Server: 9756\\ 
\hline
\multirow{2}{4em}{1,000,000} & Client:211 & Client 96702\\ 
& Server:196524 & Server: 102185 \\ 
\hline
\end{tabular}
\end{center}
\label{table:server}
\end{table*}
\fi
%%%%%%%%%%%%%%%%%%%
\begin{table*}[ht]
\caption{Offline and online run-times for PSI-based protocol (in milliseconds) at the client (individual) and server (health authority) with varying size for client's local contact list. Size of server's database is set to $100,000$.}
\begin{center}
\begin{tabular}{ |c|c|c| } 
\hline
Size of client's contact list & Offline Time (ms) & Online Time (ms) \\
\hline
\multirow{2}{4em}{10} & Client: 2.7 & Client: 9852.95 \\ 
& Server: 20012.75 & Server: 9560.8 \\ 
\hline
\multirow{2}{4em}{100} & Client: 22.6 & Client: 9854.2 \\ 
& Server: 20218.1& Server: 9968.65 \\ 
\hline
\multirow{2}{4em}{1,000} & Client: 202.3 & Client: 9817.5 \\ 
& Server: 20448 & Server: 9979.75 \\ 
\hline
\multirow{2}{4em}{10,000} & Client: 1990.4  & Client: 9787.45  \\ 
& Server: 20246.45 & Server: 9970.65\\ 
\hline
\end{tabular}
\end{center}
\label{table:client}
\end{table*}

\iffalse
\begin{table*}[ht]
\caption{Offline and online run-times (in milliseconds) at the client (individual) and server (health authority) with varying size for client's local contact list. Size of server's database is set to $100,000$.}
\begin{center}
\begin{tabular}{ |c|c|c| } 
\hline
Size of client's contact list & Offline Time (ms) & Online Time (ms) \\
\hline
\multirow{2}{4em}{10} & Client:4 & Client:10391 \\ 
& Server:20473 & Server: 10471 \\ 
\hline
\multirow{2}{4em}{100} & Client: 23 & Client:10014 \\ 
& Server:20042 & Server: 10317 \\ 
\hline
\multirow{2}{4em}{1,000} & Client: 229 & Client: 10458 \\ 
& Server: 21544 & Server: 10359\\ 
\hline
\multirow{2}{4em}{10,000} & Client: 1955 & Client: 9772 \\ 
& Server: 19836 & Server: 10552\\ 
\hline
\end{tabular}
\end{center}
\label{table:client}
\end{table*}
\fi

%%%%%%%%%%%%%%%%%%%%%%%%%APSI%%%%%%%%%%%%%%%%%%%%%%%%%%%%%%%%%

\begin{table*}[ht]
\caption{Offline and online run-times for APSI-based protocol (in milliseconds) at the client (individual) and server (health authority) with varying size for server's database. Size of client's contact list is set to $1,000$.}
\begin{center}
\begin{tabular}{ |c|c|c| } 
\hline
Size of server's database & Offline Time (ms) & Online Time (ms) \\
\hline
\multirow{2}{4em}{1,000} & Client: 1082.25 & Client: 30.55 \\ 
& Server: 508.6  & Server: 512.55 \\ 
\hline
\multirow{2}{4em}{10,000} & Client: 1116.15 & Client: 25.3 \\ 
& Server: 4727.15  & Server: 483.85  \\ 
\hline
\multirow{2}{4em}{100,000} & Client: 1233.15  & Client: 49.25 \\ 
& Server: 46202.1  & Server: 496 \\ 
\hline
\multirow{2}{4em}{1,000,000} & Client: 1019.45  & Client: 118 \\ 
& Server: 454721.25 & Server: 509.55\\ 
\hline
\end{tabular}
\end{center}
\label{table:serverAPSI}
\end{table*}
%%%%%%%%%%%%%%%%%%%%%%%%%%%%%%%%%%%%%%%%%%%%%%%%%

\begin{table*}[ht]
\caption{Offline and online run-times for APSI-based protocol (in milliseconds) at the client (individual) and server (health authority) with varying size for client's local contact list. Size of server's database is set to $100,000$.}
\begin{center}
\begin{tabular}{ |c|c|c| } 
\hline
Size of client's contact list & Offline Time (ms) & Online Time (ms) \\
\hline
\multirow{2}{4em}{10} & Client: 106.7  & Client: 15.75  \\ 
& Server: 46675.4 & Server: 6.45 \\ 
\hline
\multirow{2}{4em}{100} & Client: 302.4 & Client: 22.75  \\ 
& Server: 46646 & Server: 61.4\\ 
\hline
\multirow{2}{4em}{1,000} & Client: 1151.05 & Client: 46.7  \\ 
& Server: 45700.35 & Server: 480.95 \\ 
\hline
\multirow{2}{4em}{10,000} & Client: 9507.8  & Client: 170.3  \\ 
& Server: 46946.7  & Server: 4776.35 \\ 
\hline
\end{tabular}
\end{center}
\label{table:clientAPSI}
\end{table*}

\section{Discussion}\label{sec:discussion}

In this section, we first discuss an extension of the proposed algorithm, in which we prevent a malicious individual from modifying their local contact history. Then, we discuss about potential additional features of the proposed technique. 

\subsection{APSI-Based Protocol Against a Malicious Individual}\label{sec:APSIbased}

As discussed, a malicious individual may tamper with their local contact history to learn the diagnosis of some target individuals. To prevent this, one option is to record each contact along with a corresponding digital signature from a centralized authority, such as the telecom operator (as discussed in Section~\ref{sec:contact}). Here, we describe how such signatures can be used when computing the intersection between the individual and the health authority's database. 

For this, we propose using the authorized private set intersection (APSI) protocol, in which the entries in the local contact history of an individual are digitally signed by a centralized authority and the validity of these signatures are verified by the health authority during the protocol. We discuss the details of this APSI-based protocol in the following.

\begin{figure}[ht]
\centering
		\includegraphics[width=0.80\columnwidth]{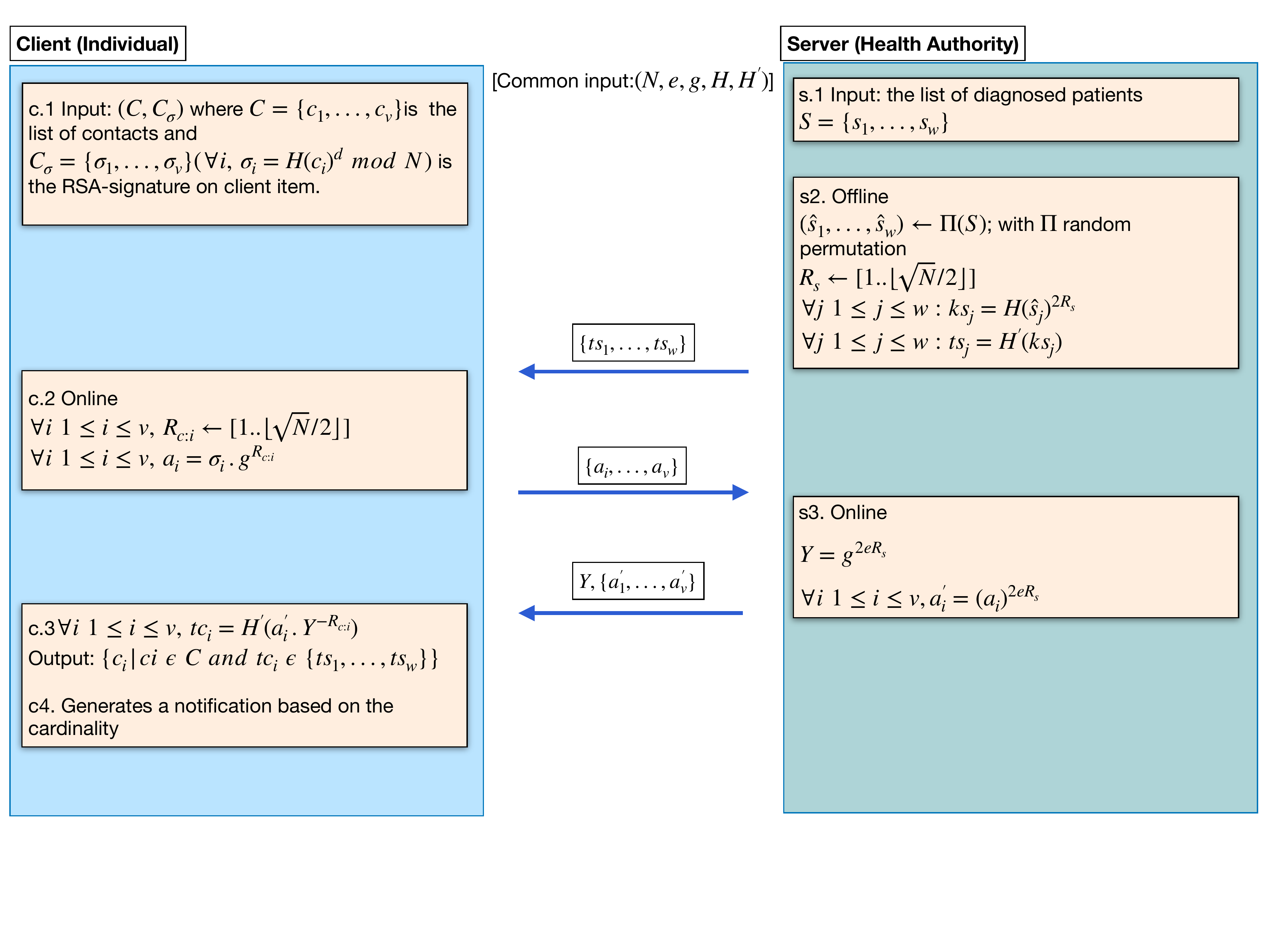}
		%\vspace{-5mm}
		\caption{Details of the APSI based protocol between an individual (client) and the health authority (server).}
		\label{fig:APSI}
		%\vspace{-3mm}
\end{figure}

Figure~\ref{fig:APSI} illustrates the details of APSI-based protocol between an individual (client) and the health authority (server) in the semi-honest setting. First, a common input $(N,e,g,H,H')$ is determined for the protocol. $N=pq$ is the RSA modulus, where $p$ and $q$ are safe primes. $e$ is the public exponent. $g$ is a random element in $\mathbb{Z}_N^*$. Also, $H$ and $H'$ are the hash functions, modeled as random oracles. All computations are done in mod $N$. Both server and the client have the same input sets as in PSI-CA protocol. In addition to these sets, client also has a list of RSA signatures $(\sigma_i)$-s, where $\sigma_i= H(c_i)^d~ mod ~N$ (steps c.1 and s.1 in the figure).

As an offline step, server permutes its input list and masks its input. For this, the server first applies the function $H(.)$ over the list of its input elements and exponentiates each element with randomness $2R_s$. Then, hash function $H'$ is applied to the list to obtain $ts_j$ values (step s.2 in the figure). These values are then sent to the client.

As an online step, client masks $\sigma_i$-s by multiplying them with $g^{R_{c:i}}$, where ${R_{c:i}}$ is the randomness (step c.2 in the figure). Client sends the resulting list that contains $a_i$ values to the server.

At server's online step, server computes $Y=g^{2eR_s}$. Server exponentiates $a_i$ values with $2eR_s$ and obtains the list of $a^{'}_i$-s (step s.3 in the figure). Server sends $Y$ and $a^{'}_i$-s to the client.

At the last step, client obtains the $tc_i$ values by applying the function $H'(.)$ over the product of  $a^{'}_i$ and  $Y^{-R_{c:i}}$. In order to get the size of the intersection, client finds the matches between the lists of $ts_j$-s and $tc_i$-s (step c.3 in the figure). At step c.4, a notification is generated based on the output of APSI.

We show the evaluation results of APSI-based solution in Tables~\ref{table:serverAPSI} and~\ref{table:clientAPSI}. Similar to the setting for the PSI-based solution in Section~\ref{sec:evaluation}, we ran our experiments on macOS High Sierra, 2.3 GHz Intel Core i5, 8GB RAM, and 256GB hard disk. We used MD5 as the hash function to hash the UIDs of individuals in the local contact lists and in the health authority's database. We used the implementation of APSI in~\cite{baldi2011countering}, in which $q$ and $p$ are $160$ and $1024$ bits, respectively. We ran each experiment for $20$ times and reported the average. We set the size of client's local contact list to $1,000$ and vary the size of server's database in Table~\ref{table:serverAPSI}. In Table~\ref{table:clientAPSI}, we set the size of server's database to $100,000$ and vary the size of client's local contact list.

We observed that APSI-based solution's online phase can be efficiently completed by the parties, even when the parties have significantly large input sizes. Similar to the PSI-based solution, the server does not need to run the offline part of the algorithm for each client separately, as it can use the same offline computation during its interaction with every client. Similarly, a client can perform its offline step using its local contact list.

%Based on the evaluation results reported in~\cite{baldi2011countering}, when the database size of the health authority is around $3$ million and the local contact history of the individual is $6$: (i)  the offline part of the protocol takes around 200 minutes for the health authority (and in our protocol, this part can be done once for all the individuals) and offline time for an individual takes negligible time, (ii) online part of the protocol takes around 2 milliseconds both for the health authority and the individual, (iii) computation cost of the online part scales linearly with the size of the individual's local contact history, and (iv) communication costs for the individual and the health authority are around $750$B and $4$GB, respectively. It is worth noting that the communication cost of the individual scales linearly with the size of the individual's local contact history and the health authority can send the same message to all individuals (it does not need to send a separate $4$GB data to each individual, and hence the communication cost of the health authority can be optimized). 
 
\subsection{Additional Features of the Proposed Technique}

The proposed technique can also be used to provide real-time whereabouts of diagnosed individuals. Having this information, healthy individuals would know which locations to stay away at any given time. In fact, such an approach has been used by some countries during the recent Coronavirus disease pandemic~\cite{verge}. However, we believe such a usage of the system may cause social chaos and it may also result in deanonymization of diagnosed patients' identities (even if the information is shared in a differentially-private way~\cite{dwork2014algorithmic}). Therefore, we prefer not to include this functionality in the proposed system.  

\subsection{Mitigation Against the Considered Attacks}\label{sec:mitigation}

Here, we discuss the possible mitigation techniques for threats explained in Section~\ref{sec:threat}.
\begin{enumerate}
    \item  In order to prevent the attack, in which a curious user tries to infer contact information or diagnosis belonging to a target user, a user should not be given read/write permission to the stored contact list on the device (or the contact list should be kept encrypted in the device). Doing so would prevent an attacker target specific victims by modifying its contact list.
    \item To avoid a curious user including fake contacts in the contact list, a digital signature can be included during the contact. The details of this solution are discussed in Section~\ref{sec:APSIbased}. %EA: can you give more details here? or refer to the section where we discussed this?
    \item The attack, in which a server colludes with a curious user to infer contact information or diagnosis belonging to another target user can be addressed as follows: If a curious user can control the contact history, then by colluding with the server, the user can learn the diagnosed people in its contact list. However, if the user cannot control the contact list, the only instance the user can learn whether a person is diagnosed or not is the case where the user has only one person in the contact list. If the user was intentionally in contact with only one person, then the user would know whether that person is positive or not. If the user receives the message that indicates a contact with a diagnosed person, then the user will find out that the person on the contact list is diagnosed with the disease.

\end{enumerate}

\section{Conclusion}\label{sec:conclusion}

In this paper, we have proposed a privacy-preserving technique to control the spread of a virus in a population. The proposed technique is based on private set intersection between physical contact histories of individuals (that are recorded using smart phones) and a centralized database (run by a health authority) that keeps the identities of the positive diagnosed patients for the disease. We have shown that individuals can receive warning messages indicating their previous contacts with a positive diagnosed patient as a result of the proposed technique. While doing so, neither of the parties that involve in the protocol obtain any sensitive information about each other. We believe that the proposed scheme can efficiently help countries control the spread of a virus in a privacy-preserving way, without violating privacy of their citizens.

%  \bibliographystyle{splncs04}
%  \bibliography{references}
\end{document}